\documentclass[pra,aps,amsmath,amssymb,superscriptaddress,twocolumn,longbibliography]{revtex4-1}
\usepackage{graphicx,multirow}
\usepackage{color,outlines}
\usepackage[sc,osf]{mathpazo}
\usepackage{hyperref}
\usepackage{physics}
\usepackage{outlines}

\begin{document}

\title{Complexity growth for one-dimensional free-fermionic lattice models}

\author{S. Aravinda } 
\email{aravinda@iittp.ac.in}
\author{Ranjan Modak} 
\email{ranjan@iittp.ac.in} 

\affiliation{Department of Physics, Indian Institute of Technology Tirupati, Tirupati, India~517619} 
\begin{abstract}
   Complexity plays a very important part in quantum computing and simulation where it acts as a measure of the minimal number of gates that are required to implement a unitary circuit.  We study the lower bound  of the complexity [Eisert, Phys. Rev. Lett. 127, 020501 (2021)] for the unitary dynamics of the one-dimensional lattice models of non-interacting fermions.  We find analytically using quasiparticle formalism, the bound grows linearly in time and followed by a saturation for short-ranged tight-binding Hamiltonians. We  show numerical evidence that for
   an initial Neel state  the bound is maximum  for tight-binding Hamiltonians as well as for the long-range hopping models. However, the increase of the bound is sub-linear in time for the later, in contrast to the linear growth observed for short-range models. The upper bound of the complexity in non-interacting fermionic lattice models is calculated, which grows linearly in time even beyond the saturation time of the lower bound, and  finally, it also saturates.
\end{abstract}

\maketitle

\section{Introduction} 

 The complexity has been playing a dominant role in a wide variety of studies in both  humanities and natural sciences~\cite{peliti1988measures,erdi2008complexity,nicolis1989exploring,goldreich2008computational,colander2014complexity,rosser1999complexities,day1994complex,rosser1999complexities,mainzer2004thinking}
. Historically, complexity emerged to understand the time a computer takes to complete a task, and it later engulfed many research areas. Distinct notions of complexity have been defined in various research fields in which it quantifies how {\it hard} the given task is. The possibility of computation of any
computable function by the computing machine, as en-
visaged by the Church-Turing thesis makes computational complexity a well-defined quantification of complexity. It quantifies complexity as a minimum number of time steps required to complete the given task in a Turing machine. The ease (or hardness) of a task depends on how the complexity varies with the size of the problem~\cite{arora2009computational}. 
The complexity of classical dynamics, quantified as the algorithmic complexity~\cite{chaitin1990information} and  the Kolmogorov complexity, can then be characterized using computational complexity~\cite{
li2008introduction,wallace1999minimum,shen2022kolmogorov}. 

Nature at its depth is quantum mechanical. Historically, after the success of a few elementary algorithms in quantum computation~\cite{deutsch1989quantum,deutsch1992rapid,shor1994algorithms,simon1997power}, the notions of quantum computational complexity  classes were defined~\cite{bernstein1993quantum}. The universality of a set of quantum gates for simulating any unitary operator was proved \cite{kitaev2002classical}. Quantum complexity theoretic tools were used to understand the possibility of estimating the quantities of quantum many-body systems from the quantum computer.  It is known that estimating the ground state energy of  the Hubbard model~\cite{schuch2009computational,childs2014bose}, the Heisenberg model~\cite{cubitt2016complexity} and the nearest-neighbor interactions in one and  two dimensional systems~\cite{aharonov2009power,hallgren2013local,oliveira2005complexity} are Quantum Merlin Arthur (QMA) complete. Many related  results on the computational complexity of the transverse field Ising model~\cite{bravyi2017complexity} is shown. The computational undecidability (uncomputability)~\cite{rozenberg1995cornerstones} is also proven for various quantities in  quantum many-body systems~\cite{bausch2020undecidability,bausch2021uncomputability,cubitt2022undecidability,watson2022uncomputably,shiraishi2021undecidability}.  The complexity in the form of possibility of efficient simulation in noisy intermediate-scale quantum (NISQ) computers and the quantum supremacy is also studied for various quantum many-body systems~\cite{maskara2022complexity,shtanko2021complexity,ehrenberg2022simulation}.

At this juncture, we would like to ask how difficult  the many-body quantum {\it dynamics} from the perspective of quantum computation. For example, consider a pure product many particle initial state $\ket{\Psi}_\text{in}$ is evolved to some other final state $\ket{\Psi}_\text{fin} = U \ket{\Psi}_\text{in} $. The complexity of the process is essentially the complexity of implementing the unitary operator $U$. 
The complexity is quantified in terms of gate complexity which is defined as the number of universal gates minimized over all the universal gate sets required to simulate the unitary dynamics. As there can be many universal gate sets, minimizing overall universal gate sets is tedious, leading to difficulty in calculating gate complexity. To counter this, Nielsen (with others) has developed techniques to find the lower bounds on the gate complexity~\cite{nielsen2005geometric, dowling2006geometry, nielsen2006quantum}.    

Recently, Eisert \cite{eisert2021entangling} showed the lower bound on gate complexity that depends on the entanglement produced by the unitary operator acting on the pure product states. However, note that entanglement and gate complexity are different. The swap gate does not produce any entanglement, yet the complexity of producing the swap gate is nonzero. In this work, we would like to study the gate complexity growth for many-body dynamics. Complexity growth has been playing an interesting role in understanding the quantum gravity models (See review~\cite{chapman2022quantum}). The conjecture~\cite{brown2018second}  on the growth of complexity with random unitary operators has been proved recently~\cite{haferkamp2022linear}. Our main goal of this paper is to study the lower bound of the complexity growth for Hamiltonian systems. 
To explore this possibility here we consider long-range power-law hopping models of non-interacting fermions. 
Recent advancements in  experiments with 
atomic, molecular, and optical systems (AMO) \cite{exp1,exp2}, power-law hopping with tunable exponent $0 <\alpha < 3$ can
be realized in laser-driven cold atom setup \cite{exp3,exp4}. 
However, we like to emphasize that these experimentally realizable models are for hard-core bosons. 
In the dilute limit, where the role of interaction is week, potentially our results for non-interacting fermions can still be relevant in the context of AMO experiments.
The dipolar ($\alpha=3$) and van-der-Waals ($\alpha=6$)
couplings  also  have been experimentally realized in the case of neutral atoms and Rydberg atoms \cite{exp5,exp6,exp7,exp8}. 
Hence, there has been a plethora of work on long-range models in last 
few years~\cite{arti1,arti2,modak.prr,chatterjee2022one,Burin2015,Burin2015b,modak.20}.
We also investigate the tight-binding limit i.e. $\alpha \to \infty$, where the entanglement growth can be well understood within the quasiparticle
frame work~\cite{calabrese2005evolution,calabrese_cardy_06}. 
Quasiparticle description can be extended for other interacting integrable models as well, where a statistical description of local properties of the steady state is possible in terms of a Generalized Gibbs Ensemble (GGE)~\cite{vidmar2016generalized} in contrast to the generic non-integrable models, where the stationary behaviour of local and quasilocal observables is described by the Gibbs (thermal) ensemble~\cite{srednicki1994chaos,deutsch1991quantum}.

The paper is organized as follows. In Sec.II we introduce geometrically local circuit cost, and in Sec III we discuss the model. Next, we discuss the results in Sec. IV which consists of two parts, first 
the tight-binding model and then the long-range model. Finally, in Sec. V we summarize our results and conclude.

\section{ Geometrically local circuit cost: }
The quantum circuit complexity of any unitary operator $U$ is the number of basic gates minimized over  universal gates. As the single particle gates don't constitute entanglement generation and from the quantum computation perspective, it is easy to operate (within the error bounds), the circuit complexity counts only nonlocal gates.  If the quantum circuit simulating the unitary dynamics contains only operators acting on single and two-particle systems, the quantum complexity is termed as {\it geometrically local circuit cost} and denoted as $C_g(U)$. 

Consider a lattice of $L$ particles with each dimension $d$.  An unitary dynamics $U \in SU(d^L)$ can be generated by the Hamiltonian $H(k)$ as 
\begin{equation}
    U = \mathcal{K} \exp (-i \int_0^1  dk ~ H(k)).
    \label{eq:udec}
\end{equation}
 $H(k)$ can be expressed with the traceless Hermitian operators $h_1,h_2, \cdots, h_J$ acting on two-particle with norm $||h_j|| =1,~\forall j$ as 
\begin{equation}
    H(k) = \sum_{j=1}^J  y_j(k) h_j, 
    \label{eq:decomp}
\end{equation}
with $y_j : [0,1] \rightarrow \mathbb{R}$ is the cost function. The geometrically local circuit cost $C_g(U)$ of an unitary operator $U$ is 
\begin{equation}
    C_g(U) := \inf \int_0^1 \sum_{j=1}^J |y_j(k)| dk,
    \label{infinum}
\end{equation}
where the infimum is taken over all the continuous function $y_j$ satisfying Eq.~(\ref{eq:udec}) and Eq.~(\ref{eq:decomp}). 

Note that while it seems that for any Hamiltonian $H(k)$ that can be written as $H(k) = \sum_{j=1}^J  y_j(k) h_j$, the above integration can be computed in straight forward manner, and hence obtaining $C_g(U)$ should be an easy task. We like to emphasize an infimum needs to be computed over all such representations of $H(k)$, that makes computing  exact geometrically local circuit cost for any Hamiltonian an extremely tedious job. Hence, in this work we are interested in calculating the lower bound.
The geometrically local circuit cost $C_g(U)$ is lower bounded by the entanglement generation by the unitary operator $U$ as~\cite{eisert2021entangling} 
\begin{equation}
    C_g (U) \geq \frac{1}{c \log (d)} \sum_{i=1}^{L-1} \mathbb{E} (U \ketbra{\phi} U^\dagger : i), 
    \label{eq:gcom} 
\end{equation} 
where 
\begin{equation}
\mathbb{E} (U \ketbra{\phi} U^\dagger : i) := S(\tr_B (\rho)) 
\end{equation}
is the entanglement entropy over the bipartision $A = \{1,2, \cdots i\}$ and $B = \{ i+1, \cdots L\}$, $c>0$, and $\ket{\phi}$ is pure product state.  The constant $c$ emerges from lower bounding the entangling rate introduced in Ref.~\cite{bravyi2007upper}. We consider $c=2$ as optimal value  obtained  in Ref.~\cite{marien2016entanglement,audenaert2014quantum}. The von Neumann entropy $S(\rho)$ is $S(\rho) = -\Tr \rho \ln \rho$. 

We are interested in studying the quantity that bounds the  $C_g(U)$  and we call it as {\it geometric entanglement capacity} (GEC) of a unitary operator $U$ and defined as 
\begin{equation}
   E_g(U) :=  \frac{1}{ \log (d)} \sum_{i=1}^{L-1} \mathbb{E} (U \ketbra{\phi} U^\dagger : i),
   \label{eqnGEC}
\end{equation}

As the $\ket{\phi}$ is some pure product state it might so happen that if the choice of the initial pure product state is inappropriate, it does not capture the actual complexity of an operator. Hence we define the following quantity as maximum GEC by maximing over all pure product states $\ket{\phi}$, and we call the upper bound of maximum GEC of an unitary operator $E_g^{\max} (U)$, 
    \label{eq:maxGEC}
which takes $E^{\max}_g(U) = \frac{L^2}{4}$. 
 This can be easily seen as follows. If $L$ is even and owing to the fact that symmetry of entanglement as measured by the entanglement entropy we can write 
    \begin{equation}
        \begin{split}
         E_g^{\max}(U) & = \frac{1}{\log d} \{ 2 [ \log d + \log d^2 + \cdots + \log d^{(\frac{L}{2}-1)} ] \\ 
        & + \log d^{\frac{L}{2}} \} = \frac{L^2}{4}. 
         \end{split}
    \end{equation}      

\section{Model}
Our main focus is to investigate the lower bound of the complexity for models of free fermions. 
We study noninteracting fermions in 1D lattice of size $L$. The system is 
described by the following  long-range power-law hopping Hamiltonian, 
\begin{eqnarray}
 {H}=-\sum _{i,j\neq i}\frac{1}{|i-j|^\alpha}(\hat{c}^{\dag}_i\hat{c}^{}_{j}+\text{H.c.}) \nonumber \\
 \label{hamiltonian}
\end{eqnarray}
where $\hat{c}^{\dag}_i$ ($\hat{c}_{i}$) is the fermionic creation (annihilation) operator at site $i$, $\hat{n}_i=\hat{c}^{\dag}_i\hat{c}_{i}$ is the number operator. 
While the Hamiltonian $H$ is a long-range hopping model of free fermions, in the limit $\alpha\to \infty$, 
the Hamiltonian is the same as tight-binding model with nearest-neighbor hopping, and it reads as, 
\begin{eqnarray}
H=-\sum_{i}\hat{c}_{i}^{\dag}\hat{c}_{i+1}+\text{H.c}.
\label{tb}
\end{eqnarray}
  The Hamiltonian ~\eqref{tb} can be diagonalized in the momentum 
basis and  can be written as, 
\begin{eqnarray}
H=\sum_{k} \epsilon_k \hat{c}_{k}^{\dag}\hat{c}_{k}
\label{tb1}
\end{eqnarray}
where, $\epsilon _k=-2\cos k$. $\hat{c}_{k}^{\dag}$ and $\hat{c}_{k}$ are fermionic creation and annihilation operators in momentum basis. All the results presented in the paper are for half-filling cases, i.e. number of fermions $N=L/2$ and $L$ is chosen to be an even number. However, we have have considered other filling fractions and found qualitatively similar results.

\section{Results}
In order to obtain GEC, we prepare the system in a non-equilibrium pure product state $|\phi\rangle$ 
and then let it evolve under the unitary dynamics governed by a Hamiltonian $H$. The calculation of GEC 
involves the bipartite von Neumann (entanglement) entropy, which is given by $S=-\textrm{Tr}\rho_A\ln\rho_A$, 
where the reduced density matrix  $\rho_A$ is defined as $\rho_A\equiv\textrm{Tr}_B|\phi(t)\rangle\langle\phi(t)|$. The trace is over the degrees of freedom of the complement  $B$ (which is of the size $L-\ell$) of $A$ (which is of the size $\ell)$, 
and $|\phi(t)\rangle= e^{-i H t} |\phi\rangle$ is the time-dependent state of the system.

\begin{figure}
    \centering
    \includegraphics[width=0.46\textwidth]{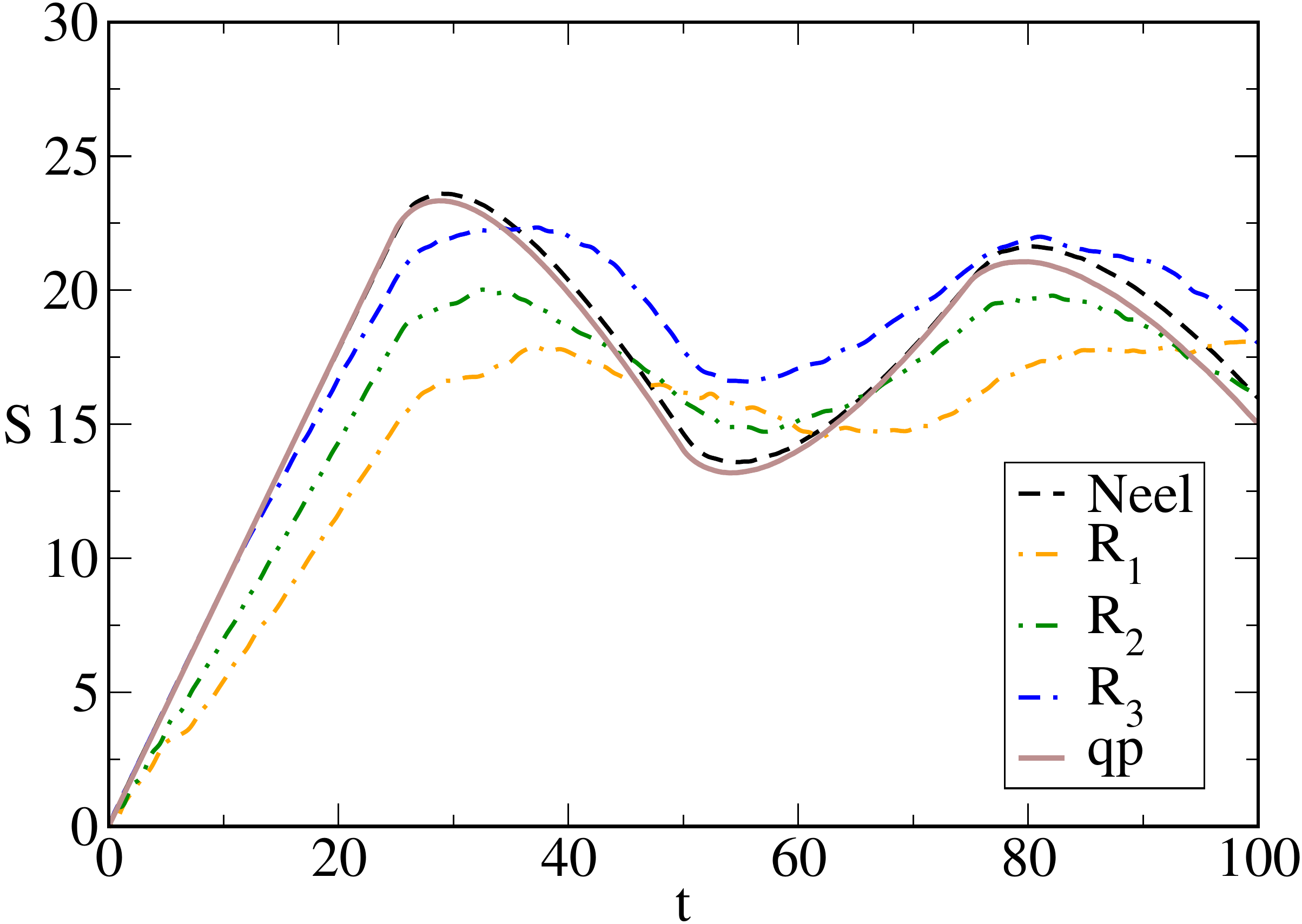} 
        \includegraphics[width=0.46\textwidth]{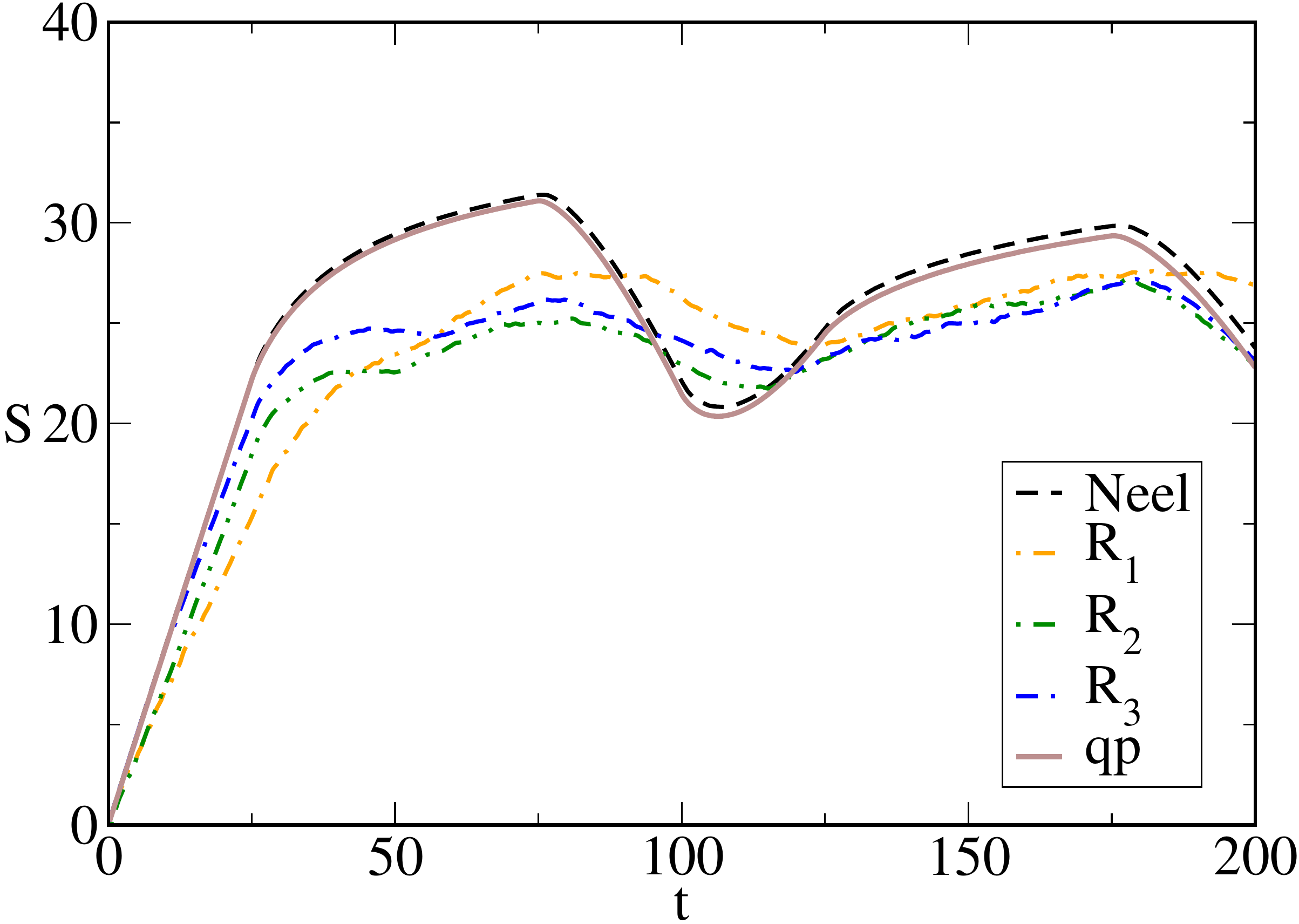}

    \caption{Entanglement dynamics for the Hamiltonian ~\eqref{tb} and for different initial states. Upper panel results are for  (a)$\ell=50$ and $L=100$, lower panel results correspond to (b)$\ell=50$ and $L=200$. The solid line corresponds to quasiparticle prediction for the Neel state.}
    \label{fig1}
\end{figure}

\begin{figure}
    \centering
    \includegraphics[width=0.46\textwidth]{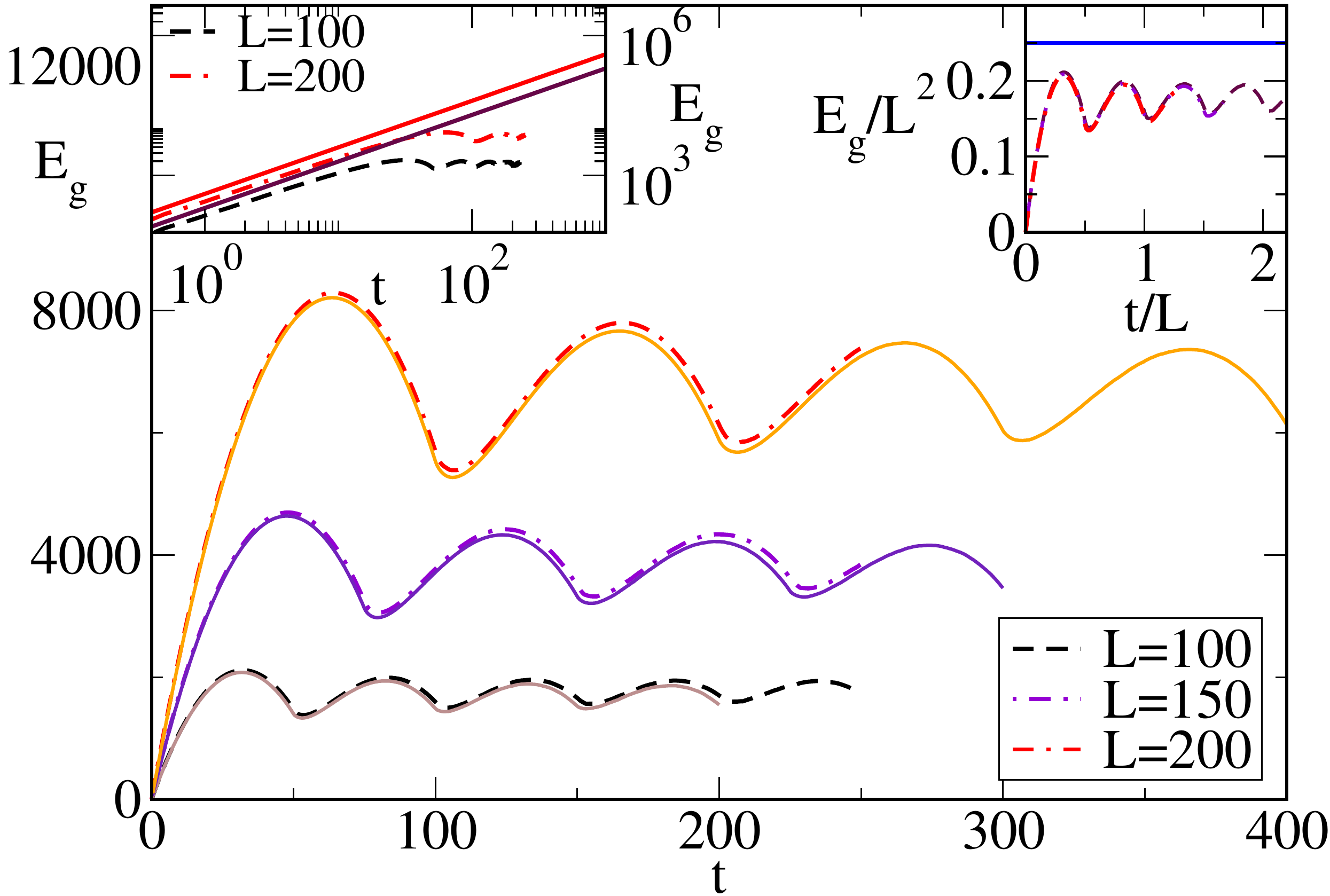} 

    \caption{Variation of $E_g$ for a tight-binding model with time for initial Neel state and for different values of $L$. 
    Solid lines correspond to quasiparticle prediction.
    Top Right Inset shows data collapse in $E_g/L^2$ vs $t/L$ plot. The solid blue line corresponds to the upper bound of $E_g/L^2$. The top left inset shows the variation of $E_g$ with time and the upper bound of the $E_g$ in solid lines.}
    \label{fig2}
\end{figure}

\subsection{Tight-binding model: Numerical results vs Quasiparticle picture}

First, we investigate the tight-binding model, and our aim is to find out a pure product state  $|\phi\rangle$ that maximizes 
GEC (Eqn.~\ref{eq:maxGEC}). Given the expression of GEC essentially involves von Neumann entanglement entropy $S$, we investigate the growth of $S$ with time for different initial product states in Fig.~\ref{fig1}. We find numerical evidence that the entanglement growth for the Neel state i.e. $|\phi\rangle=\prod_{i=1}^{L/2}\hat{c}_{i}^{\dag}|0\rangle$  is the maximum in the short time regime. We compare the entanglement dynamics for Neel state with different possible random initial states, among which we report three such  
states in our paper which are denoted as $R_1$ ($L/2$ fermions are completely randomly distributed over the lattice), $R_2$ (in 1st $50$ sites fermions are occupied in alternative sites and the rest of the fermions are completely randomly distributed over rest of the lattice), and $R_3$ (in 1st $64$ sites fermions are occupied in alternative sites and the rest of the fermions are completely randomly distributed over rest of the lattice). In all three cases, while we report our results for three individual uniformly  random realizations, but in each cases we check our results for $1000$ such random realizations and find that the growth of GEC is slower than the Neel state.

On the other hand, the out-of-equilibrium dynamics of $S$ after a quantum quench in finite-size integrable systems can be described by the quasiparticle picture. 
According to this picture, the initial state acts as a source  of quasiparticle excitations which are produced in pairs and uniformly in space. 
After being created, the quasiparticles move ballistically through the system with opposite velocities. 
Only quasiparticles created at the same point in space are entangled and while they move far apart  carrying entanglement and correlation in the system. 
A pair contributes to the entanglement entropy at time $t$ only if one particle of the pair is in $A$  and its partner in $B$.
Keeping track of the linear trajectories of the particles, it is easy to show~\cite{calabrese_cardy_06,alba2017entanglement} 
\begin{equation}
\label{semi-cl}
S(t)= \sum_n\Big[ 2t\!\!\!\!\!\!\int\limits_{\!2|v_n|t<\ell}\!\!\!\!\!\!
\frac{dk}{2\pi} v_n(k)s_n(k)+\ell\!\!\!\!\!\!\int\limits_{2|v_n|t>\ell}\!\!\!\!\!\!
\frac{dk}{2\pi} s_n(k)\Big].
\end{equation}
Here the sum is over the species of particles $n$ whose number depends on the model, $k$ represents their quasimomentum (rapidity), 
$v_n(k)$ is their velocity, and $s_n(k)$ their contribution to the entanglement entropy. 
(For the tight-binding model there is a single species of quasiparticle, hence we omit the sum over $n$ and the subscripts). 
The quasiparticle prediction~\eqref{semi-cl}  for the entanglement entropy holds true in the space-time 
scaling limit, i.e. $t,\ell\to \infty$ with the ratio $t/\ell$ fixed.  
If the maximum quasiparticle velocity is $v_M$~\cite{lieb2004finite}, Eq.~\eqref{semi-cl} predicts 
that for $t\le \ell/(2v_M)$, $S_{\ell}$ grows linearly in time. 
Conversely, for $t\gg\ell/(2v_M)$, only the second term survives and the entanglement is extensive in the subsystem size, i.e., $S\propto\ell$. In order to evaluate Eq. \eqref{semi-cl}, one needs to evaluate $v_n (k)$ and $s_n(k)$. 
The former is the group velocities of the excitations around the stationary state and the latter is the thermodynamic entropy densities of the GGE~\cite{fagotti2008evolution}. 
The thermodynamic entropy can be obtained as~\cite{alba.2018},
\begin{eqnarray}
S_{\text{GGE}}=-\Tr [\rho_{\text{GGE}}\ln \rho_{\text{GGE}}]= \sum_k -\lambda_k\frac{\partial \ln Z}{\partial \lambda _k} +\ln Z   \nonumber \\
=\sum_k -n_k\ln n_k-(1-n_k) \ln (1-n_k) =\sum _k s_k, \nonumber\\
\label{s_k_tb}
\end{eqnarray}
where, $s_k=-n_k\ln n_k-(1-n_k) \ln (1-n_k)$ is identified as the entropy contribution of the 
quasiparticle with momentum $k$. On the other hand, $v_k$ can be obtained from the dispersion relation of the Hamiltonian \eqref{tb}, i.e. $v_k=d\epsilon_k/dk=2\sin k$
(the maximum quasiparticle velocity $v_M=2$, which corresponds to $k_M=\pi/2$). 
This formula is generically valid for
free fermionic models with the crucial assumption that the initial
state is writable in terms of pairs of quasiparticles. Other types of particular structures of initial states have also been considered recently, for which the effective quasiparticle velocities have been also computed ~\cite{eff_velocity_1,eff_velocity_2}.

Moreover, we want to find out the initial state for which GEC is maximum. 
It is obvious from the Eqn.~\eqref{s_k_tb} that the maximum value $s_k$ can have is $s_k=\ln 2$, which corresponds to $n_k=1/2$. The $n_k$ for the initial Neel state and as well as for all other half-filled random states $R_1$, $R_2$, and $R_3$ can be obtained from,
\begin{eqnarray}
 n_k=\langle \phi |\hat{n}_k|\phi \rangle 
=\frac{1}{L}\sum_{jl}e^{i(j-l)k}\langle \phi|\hat{c}_{j}^{\dag}\hat{c}_l |\phi \rangle .
\end{eqnarray}
where  the initial state is $|\phi \rangle$. 
 For our choice of initial state it is straight forward to see that  $n_k=\langle \phi |\hat{n}_k|\phi \rangle = \frac{1}{2}$, which implies  $s_k=-n_k\ln n_k-(1-n_k)\ln(1-n_k) =\ln2$ and the maximum value of $S$ is $\ell\ln2$. Moreover, the maximum of GEC for all such state can be easily evaluated as,
\begin{eqnarray}
    E_g&=&\frac{2}{\ln 2}\sum_{\ell=1}^{L/2-1}\ell \ln2+\frac{L}{2\ln 2}\ln 2 \nonumber \\
    &=&2(\frac{L}{2}-1)\frac{L}{2} +\frac{L}{2}=\frac{L^2}{4}, \nonumber
\end{eqnarray}
and this is the upper bound of GEC for any Unitary operator. 
While our numerical evidence in Fig.~\ref{fig1} supports our analytical finding i.e.  in the long-time limit entanglement entropy (also  GEC)
for all such initial states tend to saturate to the same value, but the short-time growth is much faster for the initial Neel state compared to other random states.
Also, fig.~\ref{fig2} shows the variation of $E_g$ with time for the Neel state
for different values of $L$. Remarkably, in the inset, we show data collapse
for $E_g/L^2$ vs $t/L$ plot. This result can also be easily understood within 
quasiparticle picture. 
Note that Eqn.~\eqref{semi-cl} works in the true space-time limit 
i.e. when $t$, $\ell \to \infty$. While here the GEC calculation 
involves summations of entanglement entropy over all the cuts that implies  
sub system size $\ell=1,2 … L/2$. Though it is expected that for relatively large $L$  and $\ell$, Eqn.~\eqref{semi-cl} works very well, but for very small values of $\ell$ (even though $L$ is large enough) there could be a discrepancy between exact entanglement entropy results and quasiparticle predictions, which has also been observed in our numerical results where in the long-time timit  $E_g/L^2$  oscillates between roughly $0.15$
 and $0.20$ which is less than $1/4$ which is essentially prediction of Eqn.~\eqref{semi-cl}.

 As pointed out before Eq.~\eqref{semi-cl} works only in the space-time scaling limit. For finite-size systems,  
one needs to modify Eq.~\eqref{semi-cl}  by carefully tracking quasi-particles 
on a circle leaving and re-entering the interval, which leads to 
the final formula \cite{Modak_2020,chapman2019complexity},
\begin{eqnarray}
S_{\ell}(t)&=&\int_{\mathrm{frac}\left(\frac{2v_{k} \, t}{ L}\right) < \frac{\ell}{L} }{\frac{dk}{2\pi} s_k L\mathrm{frac}\left(\frac{2v_{k} \, t}{ L}\right)} \nonumber \\
&+&\ell \int_{ \frac{\ell}{L}\leq \mathrm{frac}\left(\frac{2v_{k} \, t}{ L}\right) < 1-\frac{\ell}{L}}{\frac{dk}{2\pi} s_k} \nonumber \\
&+&\int_{ 1-\frac{\ell}{L}\leq\mathrm{frac}\left(\frac{2v_{k} \, t}{ L}\right) }{\frac{dk}{2\pi} s_k L\left(1-\mathrm{frac}\left(\frac{2v_{k} \, t}{ L}\right)\right)} \nonumber \\
\label{ee}
\end{eqnarray}
where $\mathrm{frac}$ denotes the fractional part, e.g.,  $\mathrm{frac}(1.42)=0.42$.   Solid lines in Fig.~\ref{fig1} and Fig.~\ref{fig2} are obtained 
using Eqn.~\eqref{ee}, it matches brilliantly with our finite-size numerical results. It is evident from 
our numerical study and also from the quasiparticle formalism, for the tight-binding model $E_g$ grows 
linearly in time and followed by a saturation. Similar behavior is conjectured in the quantum gravity studies for random unitary operators~\cite{brown2018second} and is been recently proved~\cite{haferkamp2022linear}.

It is important to emphasize that given the $E_g$ is related to the growth of a particular
measure of entanglement, this result is naturally limited to the early time regime
in which entanglement is not yet saturated. It is well-known that complexity is
expected to grow beyond this entanglement saturation time. The tight binding model (Eqn.~\eqref{tb}) on a $L$ sites lattice  can be expressed with the $L-1$ traceless Hermitian operators $h_j=\hat{c}^{\dag}_j\hat{c}_{j+1}+\hat{c}^{\dag}_{j+1}\hat{c}_{j}$ that are supported on two qubits, 
\begin{equation}
    H = \sum_{j=1}^{L-1} h_j. 
\end{equation}
Hence, an automatic consequence is $y_j=1$, $\forall j$ in Eq.~\ref{eq:decomp}. The unitary operator acting on two qubits will have the form $U= e^{-iHt}$. Using Eqn.~\eqref{infinum} an upper bound for $E_g(U)$ can be calculated and  reads as, $E^{ub}_g=c(L-1)t$.  The top left inset in Fig.~\ref{fig2} demonstrates the variation of the upper bound with time. Here we choose $c$ to be $2$ as suggested in Ref.~\cite{eisert2021entangling,marien2016entanglement,audenaert2014quantum}.
This upper bound keeps growing with time till it reaches $t = 4^L$.  The saturation is followed by the fact that any unitary operator $U$ of dimension $d=2^L$ can be simulated using utmost $4^L$ two-qubit gates~\cite{knill1995approximation,brown2022polynomial}. Remarkably, we find that the difference between the upper and lower bound of the complexity is not that significant until the lower bound gets saturated. However, the upper bound keeps on growing even beyond this point.
   
\begin{figure}
    \centering
    \includegraphics[width=0.46\textwidth]{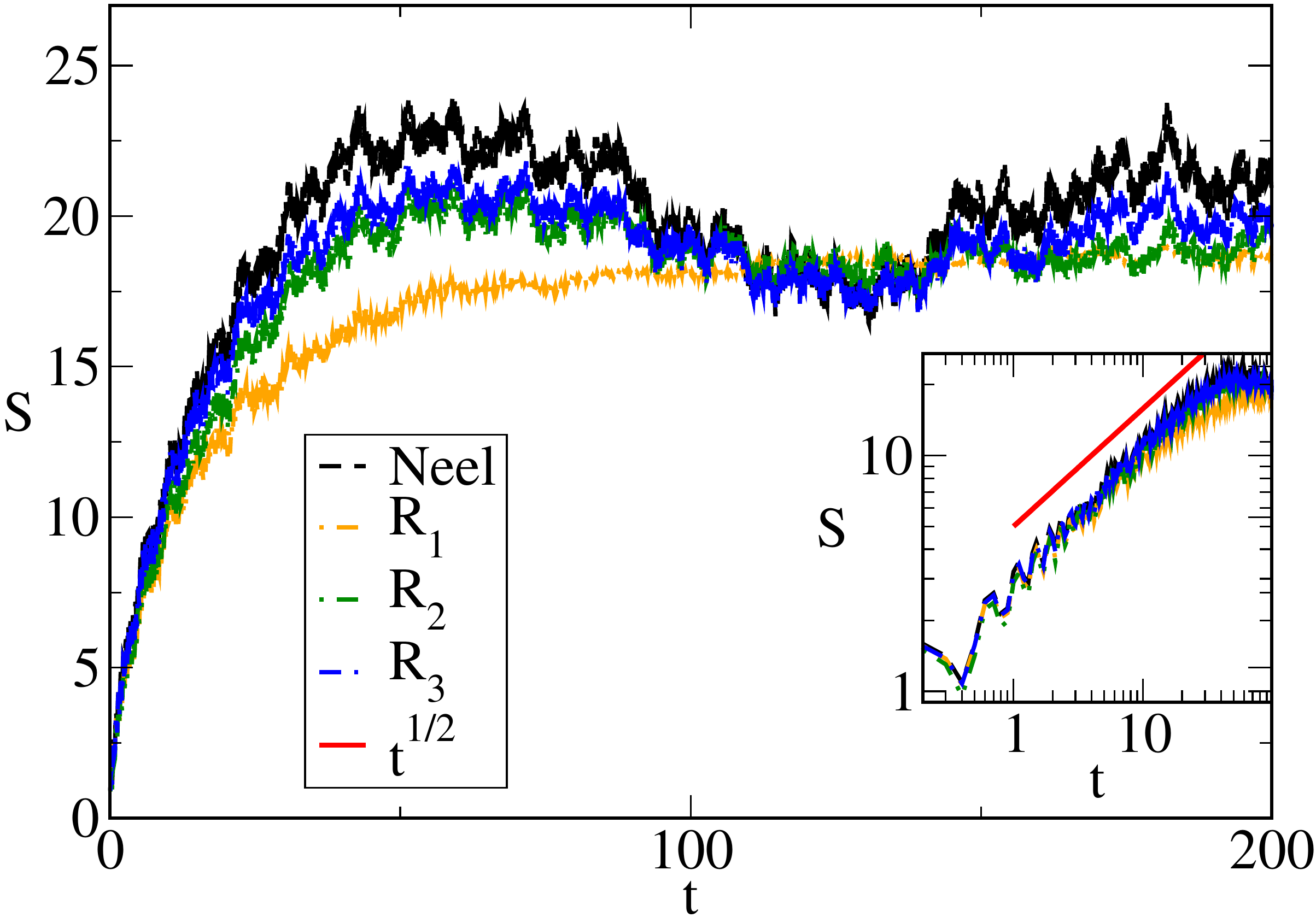} 

    \caption{Entanglement dynamics starting from different initial states for long range model $\alpha=0.5$, $L=100$, and $\ell=50$. Inset shows the growth for Neel state is $E_g\sim t^{1/2}$.}
    \label{fig3}
\end{figure}
\begin{figure}
    \centering
    \includegraphics[width=0.46\textwidth]{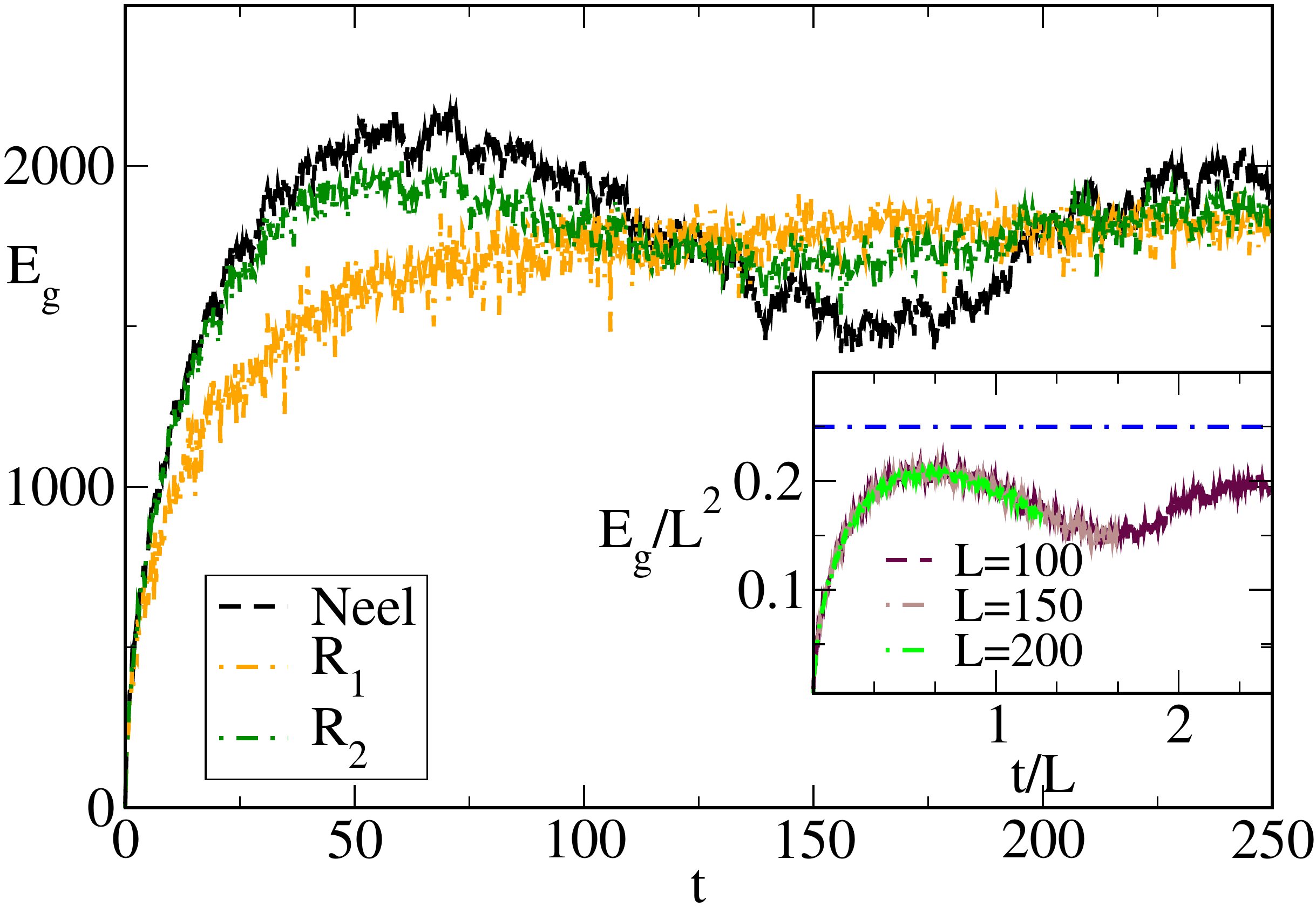} 

    \caption{Time evolution of $E_g$ for different initial states, where $\alpha=0.5$, $L=100$ and $\ell=50$. Inset shows the data collapse in $E_g/L^2 vs t/L$ plot. The blue double dashed-dot line corresponds to the upper bound of $E_g/L^2$.}
    \label{fig4}
\end{figure}
\begin{figure}
    \centering
    \includegraphics[width=0.46\textwidth]{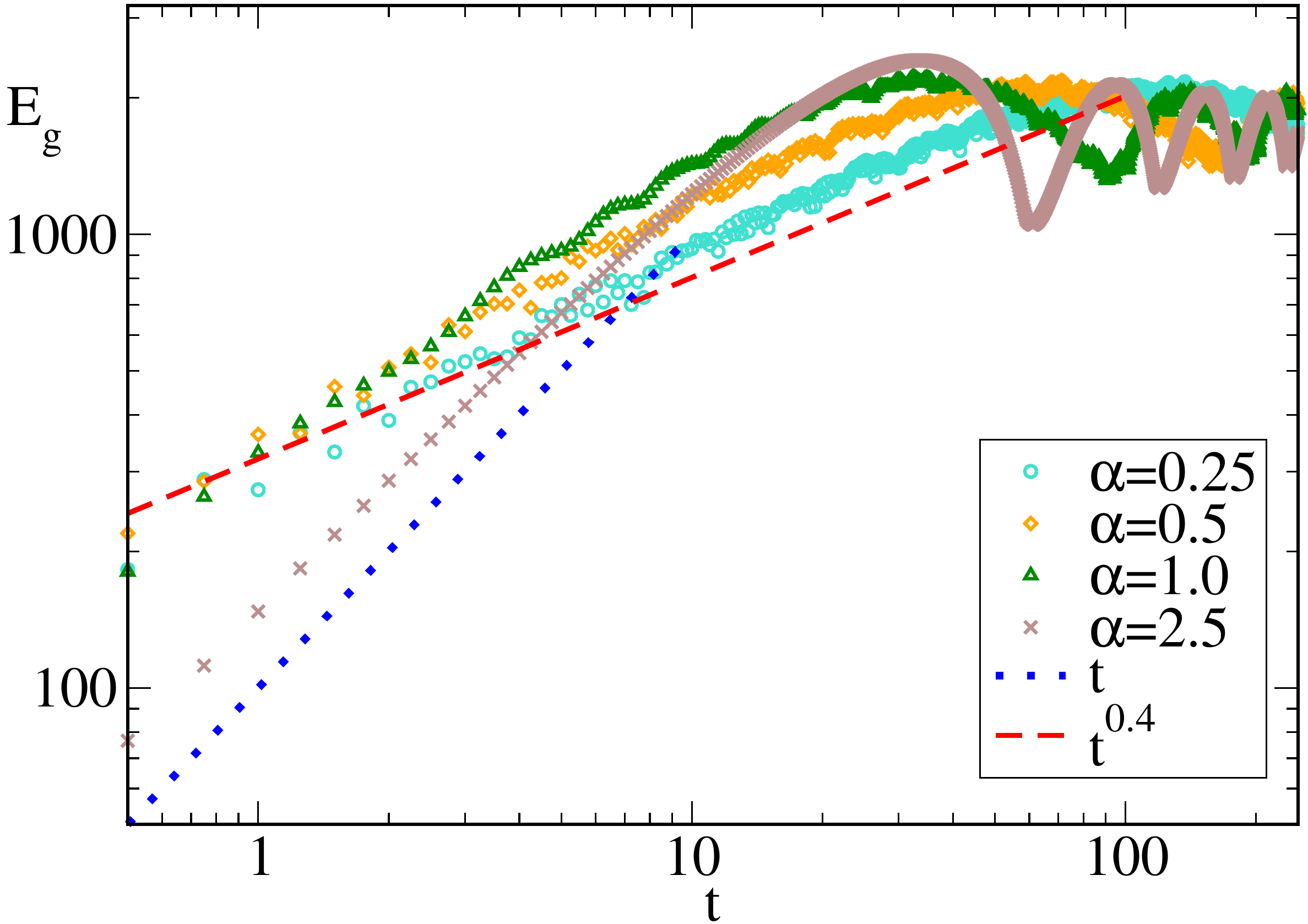} 

    \caption{Variation of GEC with time for the initial Neel state for different values of $\alpha$ and $E_g\sim t^{\gamma}$.  }
    \label{fig5}
\end{figure}
\subsection{Long-range hopping model}
 Next, we investigate the long-range hopping model. Unlike the Hamiltonian \eqref{tb}, the entanglement growth for the long-range hopping model can
not be explained within the quasiparticle picture.
Moreover, the fact that the initial state acts as a source of quasiparticle excitations which move ballistically through the system with opposite velocities, this assumption breaks down. That could lead to potential sublinear growth of entanglement instead of linear growth as predicted by Eqn.~\eqref{semi-cl}.
Figure.~\ref{fig3} shows the bipartite entanglement growth (where $\ell=L/2$) for initial Neel and other random states. Inset shows the growth is not linear in time, in contrast, $S(t)\sim t^{1/2}$ for $\alpha=0.5$. Remarkably, similar to the tight-binding Hamiltonian, even for the long-range models, the entanglement growth for the initial Neel state, turns out to be maximum compared to other states. The GEC also shows similar behaviour as shown in Fig.~\ref{fig4}. Moreover, the inset of Fig.~\ref{fig4} shows data collapse in a $E_g/L^2$ vs $t/L$ plot for different values of $L$, which was also observed for the tight-binding Hamiltonian earlier. Finally, we show in Fig.~\ref{fig5} the time evolution of $E_g$ for the initial Neel state for different values of $\alpha$. We find that for small values of $\alpha$, the growth of $E_g$ is sub-linear $E_g\sim t^{\gamma}$ in time, as $\alpha$ increases the $\gamma\to 1$.

\section{Conclusions}
In this work, we investigate the gate complexity of the one-dimensional lattice Hamiltonians of non-interacting fermions. While calculating the complexity is a tedious job, we focus on the lower bound, more specifically we study the quantity called GEC for such models. 
First, we focus on the short-ranged tight-binding model and show that the growth of GEC is linear in time and followed by a saturation. We also find numerical evidence that 
the growth is maximum for the Neel state, not only that in the thermodynamic limit for GEC saturates its upper bound i.e. $L^2/4$. We also investigate the long-range hopping model. Interestingly, we find numerical evidence that even for these models the growth of GEC is maximum for Neel state. However, given the entanglement dynamics for long-range models is sub-linear in time (in contrast to short-ranged tight-binding Hamiltonian where the growth is linear and can be explained within quasiparticle picture), followed by saturation, GEC also showed similar behaviour.  Finally, we show for both short-ranged and long-ranged models from the data collapse, GEC increases as $L^2$ 
 for a fixed $t/L$.  The entanglement saturates in a early time regime, and hence any complexity measure based on the particular entanglement measure has to be saturated quickly, but the complexity may still be growing. We found the upper bound for the non-interacting fermionic lattice model, and it increases linearly and saturates. We like to emphasize that while there have been many studies regarding entanglement growth of non-interacting systems. The validity of Eq.~\eqref{semi-cl} has been tested both analytically and numerically in free-fermion, free-boson models ~\cite{semi_1,semi_2,semi_3,alba.2018} and in many interacting integrable
models ~\cite{alba2017entanglement,piroli2019integrable,modak2019correlation,piroli2020exact}. On the other hand, the mechanism for the entanglement evolution in chaotic systems is different, not as well understood as in integrable models, nevertheless, the entanglement entropy grows linearly in early time before saturating to a value that is extensive
in subsystem size ~\cite{chaos_1,chaos_2,chaos_3,chaos_4}, exactly as in integrable systems.
 Our goal here is to study the lower bound of the circuit complexity. Though the formula for lower bound involves calculating entanglement entropy for initial product states, it also involves a maximization over all such product states, which technically is a very daunting task. One of the most important results of our work is the numerical evidence that the growth of the  GEC is the fastest for Neel state compared to any other possible random initial product states for free fermionic lattice models, hence that reduces the job of maximizing entanglement over all the product states to just calculating  GEC for only the Neel state.

 While in this paper, we restrict ourselves to the models of non-interacting fermions, GEC calculations within quasiparticle pictures can be extended for interacting integrable systems e.g. XXZ spin chain~\cite{alba2017entanglement} and Spin-1 Lai Sutherland model~\cite{piroli2019integrable,modak2019correlation,piroli2020exact}. It will be interesting to explore the complexity bounds for interacting systems both integrable and non-integrable in further studies. 
Recently  other measures of complexity is defined by various groups~\cite{girolami2021quantifying,girolami2019difficult,li2022wasserstein,balasubramanian2022quantum}, and it is interesting to see how our results can be extended with these measures.
\section{acknowledgements}
RM acknowledges the DST-Inspire fellowship by the Department of Science and Technology, Government of India, SERB start-up grant (SRG/2021/002152). SA acknowledges the start-up research grant from
SERB, Department of Science and Technology, Govt. of India (SRG/2022/000467).Authors thank the anonymous referee for extremely fruitful comments.

\bibliography{complex,ent_local}
\end{document}